\documentclass[12pt,a4paper]{article}
\begin{document}
\begin{titlepage}

\hbox{}\vspace{1in}
\begin{center}

\textsc{\Large NMR Quantum Logic Gates for Homonuclear Spin Systems.}

\vspace{1in}

\textbf{Noah Linden$^{a,b,}$\footnote{email: {\tt 
nl101@newton.cam.ac.uk}},
$\overline{\textbf E}$riks Kup\v ce$^{c,}$\footnote{email: {\tt
eriks.kupce@nmr.varian.com}},
 Ray Freeman$^{d,}$\footnote{{\rm Corresponding
author.
Fax +44-1223-336362; email: {\tt rf110@cus.cam.ac.uk}}}}

\vskip 1truecm
 
$^a$Isaac Newton Institute for Mathematical
Sciences, 20 Clarkson Road, Cambridge, CB3 0EH, UK\\
$^b$Department of Applied Mathematics and Theoretical Physics, Silver
Street,
Cambridge CB3 9EW, UK\\
$^c$Varian Ltd, 28 Manor Road, Walton-on-Thames, Surrey KT12 2QF, UK\\
$^d$Department of Chemistry, Lensfield Rd,
Cambridge CB2 1EW, UK.

\vspace{1in}

\bigskip


\end{center}
\vfill\eject
\begin{center} \textsc{\large Abstract} \end{center} \medskip
        If NMR systems are to be used as practical
quantum
computers, the number of coupled spins will need to be so large that it is
not feasible to rely on purely heteronuclear spin systems. The
implementation of a quantum logic gate imposes certain constraints on the
motion of those spins not directly involved in that gate, the so-called
``spectator" spins; they must be returned to their initial states at the 
end
of the sequence.
As a result, a homonuclear spin system where there is appreciable coupling
between every pair of spins would seem to require a refocusing scheme that 
doubles in
complexity and duration for every additional spectator spin.  Fortunately,
for the more
realistic practical case where long-range spin-spin couplings can be
neglected, simpler refocusing schemes can be devised where the overall
duration of the sequence remains constant and the number of soft pulses
increases only linearly with the number of spectator spins. These ideas 
are
tested experimentally on a six qubit system:
the six coupled protons of
inosine.
\end{titlepage}

\addtocounter{page}{2}

{\openup 5pt

\section{Introduction}

        If a system of nuclear spins is to be of real practical use as an 
NMR
quantum computer [1-13] it should consist of  tens of coupled
nuclei, otherwise the accessible algorithms will afford no real
advantage over those that can be readily implemented on a classical
electronic computer.   Unfortunately the number of suitable
(and non-radioactive)  spin-1/2 nuclei is strictly limited; the 
prime
candidates are ${}^1$H, ${}^{13}$C, ${}^{15}$N, ${}^{19}$F, ${}^{29}$Si 
and
${}^{31}$P, while chemical considerations appear to rule out the two noble
gases ${}^{3}$He and ${}^{129}$Xe, together with most of the heavy metals.
Consequently, an NMR quantum computer with a very large number of qubits 
is 
only
likely to be attainable if it includes extensive homonuclear systems of 
coupled
spins.   These spins must all form part of the same molecule, consequently
chemical bonding constraints favour the nuclei ${}^1$H, ${}^{13}$C,
${}^{19}$F.  Thus a key task is to understand how to  perform a logic 
operation on a homonuclear system, 
for
example an array of coupled ${}^{13}$C spins.  This is the difficult part 
of the
problem, whether or not heteronuclear spins are also involved, and is the 
subject of this Letter.

        A recent paper [12] has emphasized this new ``do nothing'' aspect 
of NMR quantum
computation.  For a total of $N$ homonuclear spins, all coupled to each
other, we select either one ``active" spin to perform a rotation, or two
``active" spins to evolve under their spin-spin coupling operator, leaving
$N-1$ or $N-2$ ``spectator'' spins to be returned to their initial states 
at the
end of the sequence.  Because each logic gate has an appreciable duration,
normally measured in tens of milliseconds, the ``do nothing" feature is
non-trivial,
involving the refocusing of all chemical shift and spin-spin coupling
interactions of the spectator spins, and the couplings between the active
spin(s) and all the spectators.

A fundamental constraint is that no two coupled spins should
experience simultaneous soft pulses. Although spin-spin coupling can be
neglected during a short hard radiofrequency pulse, this is not the case
for
simultaneous pulses that are selective in the frequency domain.   If the
duration of the two soft pulses is comparable with the reciprocal of the
coupling constant, undesirable antiphase magnetization and 
multiple-quantum
coherences are generated [14-17].  This so-called double-resonance 
two-spin
effect ``TSETSE'' is irreversible and interferes with the proper operation
of
the logic gate.   
This constraint dictates the form of the refocusing pulse sequence, which
can become highly complex for large numbers of spins if they are all 
coupled
together with appreciable coupling constants [12].  The soft refocusing 
pulses
eventually ``collide" in the time domain, setting a lower limit on the
duration of the logic gate.
Higher applied magnetic fields mitigate 
the
problem by increasing chemical shift differences and hence permitting
shorter (less selective) soft pulses.

        Two recent papers [18,19] describe pulse sequences that are more
efficient
in that the total number of $\pi$ pulses does not increase exponentially 
with
$N$ for a fully-coupled system of $N$ spins.  In both reports the analogy 
with
Hadamard matrices is stressed, since to refocus chemical shift or 
spin-spin
interactions requires equal periods of ``positive'' and ``negative'' 
evolution
under the $I_z$ or $2I_zS_z$ operators, and the Hadamard matrices provide 
an
elegant formulation of this requirement, suggesting the most efficient
recursive expansion procedure.  Unfortunately the proposed schemes
involve several pairs of simultaneous $\pi$ pulses and are therefore 
unsuitable for 
homonuclear systems.

	In general the design of suitable homonuclear pulse sequences can 
be based on
traditional NMR ``refocusing'' considerations, or more formally on 
patterns
derived from the Hadamard matrices.  We draw up a section of a 
conventional
Hadamard matrix in which the rows represent the different spins ($I,
\ S,\ R,\ Q$, etc.) while the columns indicate the sense of nuclear 
precession ($+$ or
$-$) in the different time segments. 
Spin-spin coupling between any two representative spins is refocused if 
the
corresponding rows are orthogonal, the characteristic
property of a Hadamard matrix.  Consequently, for a four spin system, the 
pattern:
\begin{eqnarray}
\begin{array}{ccccccccc}
                I:\quad &+      &+      &+      &+ \\
                S:\quad &+      &+      &-      &-\\
                R:\quad &+      &-      &+      &-\\
                Q:\quad &+      &-      &-      &+
\end{array}
\end{eqnarray}
using the $4\times 4$ Hadamard matrix, will ensure that spin $I$ evolves 
only according to its chemical shift, 
with
no splittings due to $S,\ R$ and $Q$, while $S,\ R$ and $Q$ have chemical 
shifts and
spin-spin splittings refocused.

	However this matrix does not satisfy the  constraint that no two 
soft $\pi$ pulses 
are
simultaneous. A possible
pattern which does satisfy the constraint is as follows: 
\begin{eqnarray}
\begin{array}{ccccccccc}
                I:\quad &+  &+  &+      &+      &+      &+      &+   
        &+\\
                S:\quad &+  &+  &+      &+      &-      &-      &-   
        &-\\
                R:\quad &+  &+  &-      &-      &-      &-      &+   
        &+\\
                Q:\quad &+  &-  &-      &+      &+      &-      &-   
        &+
\end{array}
\end{eqnarray}
This matrix involves selecting four rows from the  $8\times 8$ Hadamard 
matrix.
We see that one possible pattern of refocusing pulses forms a (1-2-4)
cascade as illustrated in Fig. 1.  

	To take a concrete example, consider first of all the case of five
homonuclear spins ($ISRQT$) all interacting with each other with 
appreciable
coupling constants.  Suppose we wish to construct a controlled-not (CNOT)
gate, which is written in terms of product operators as
\begin{eqnarray}
e^{-i{\pi\over 2}I_y}   e^{-i{\pi\over 2}(I_z +S_z)}
e^{+i{\pi\over 2}(2I_zS_z)}     e^{+i{\pi\over 2}I_y}\label{cnot}
\end{eqnarray}
where, by convention, the operators are set out in time-reversed order 
(and 
we have ignored an irrelevant overall phase).
The key step is the evolution of the $I$ and $S$ spins under the spin-spin
coupling operator $2I_zS_z$.  The overall duration of this sequence is
determined by the ``antiphase'' condition:
\begin{eqnarray}
  \tau = (2n+1)/(2J_{IS})
\end{eqnarray}
where $n$ is an integer, normally zero.  For the case of the CNOT as set 
out in (\ref{cnot}), $n$ should be even. The $R$, $Q$, and $T$ spectator 
spins
must be returned to their initial states at the end of the sequence.  
Refocusing
of the appropriate chemical shifts and spin-spin interactions is achieved
by two hard $\pi$ pulses and the (1-2-4) cascades of selective $\pi$ 
pulses
shown in
Fig. 1.  Each spin experiences an even number of $\pi$ pulses, and the 
soft
$\pi$ pulses are never applied simultaneously.

	The extension to further coupled
nuclear spins is straightforward but daunting. Additional stages are added
to the cascades, and each new stage doubles the number of time segments
and contains twice as many soft $\pi$ pulses.  Since these pulses have an
appreciable
duration, a point is eventually reached where the overall length of the 
sequence
has to be increased to accommodate so many soft pulses without overlap in
the time domain.  This would be
implemented by increasing $n$ in Eq. (4).  Herein lies the principal 
drawback
of the method, for a long sequence would be subject to appreciable
decoherence effects.  The onset of this condition is determined by the
chemical
shift dispersion of the nucleus under investigation at the field strength
of the spectrometer.   This favours ${}^{13}$C or ${}^{19}$F nuclei in the
highest possible field, because this permits the shortest soft $\pi$ 
pulses.

        Fortunately in practice spin-spin couplings are relatively local,
and in a
large spin system many of the longer-range interactions are vanishingly
small.  This puts a quite different complexion on the ``do nothing'' 
feature.
One can then find pulse patterns for the logic gate that do not incur an
exponential
increase in the number of pulses required as the number of spins is 
increased.
	
 	Consider the practical case of a system of spins disposed along a
``straight''
chain (no branching) where spin-spin couplings are limited to one-, two- 
and
three-bond interactions, neglecting the rest on the grounds that they 
would
be too weak to cause significant TSETSE effects.  This would usually be a
good approximation if we decide to study coupled  ${}^{13}$C spins in an
isotopically enriched compound. For simplicity of illustration the active
spins $IS$ have been assumed to be at the end of the chain, but identical
conclusions can be drawn for spins near the middle of a chain.

        Simultaneous $\pi$ pulses are now allowed, provided that the spins
in question
are separated by four or more chemical bonds, and provided that the two 
soft pulses are not too close in frequency [20].  This affords a dramatic
simplification and there is no longer an exponential growth in
complexity as more spins are added to the chain.  Consider a chain of nine
coupled spins ($ISRQTUVWX$).  All the relevant splittings can be refocused 
by
the application of repeated (1-2-4) cascades of soft $\pi$ pulses 
separated 
by
a stage where one spin ($U$ in this case) has no soft $\pi$ pulses at all 
(Fig.
2).   This recursive expansion can be continued indefinitely without
increasing the number of time segments in the sequence beyond sixteen.
Note that at no time are two soft $\pi$ pulses applied simultaneously to 
spins
less than four bonds apart, but that all shorter-range splittings are
refocused.  The number of soft pulses increases essentially linearly with
the total number of spins,
while the overall duration of the sequence remains constant.

	The simplification can be taken one step further by neglecting all 
except
one- and two-bond interactions.   Then a sequence made up of only eight
time segments suffices and (1-2) cascades can be employed (Fig. 3).
Simultaneous soft $\pi$ pulses are only applied to spins separated by at
least three bonds, but all shorter-range interactions are refocused.
Finally, in practical situations where all except the one-bond couplings
can be neglected, an even simpler sequence of single soft $\pi$  pulses 
can
be used.

	These ideas can be expressed in a slightly different form for   
coupled
systems of protons or fluorine nuclei.  The possible topologies comprise
the two-bond geminal couplings, three-bond vicinal couplings, and the
corresponding longer-range interactions. Since the coupling constants do
not necessarily fall off monotonically with the number of intervening
bonds, we consider them in order of decreasing magnitude, neglecting all
those below a predetermined threshold, thus avoiding the awkward and
unrealistic case where every spin interacts with every other with an
appreciable coupling constant.

	An experimental test of these proposals was carried out on six 
coupled
protons in inosine (Fig. 4) dissolved in dimethylsulfoxide-$d_6$
containing some heavy water.   The hydroxyl resonances were removed by
exchange with deuterium.  As before, the $I$ and $S$ spins were allowed to
evolve under the scalar coupling operator, while $R$, $Q$, $T$, and $U$
were passive spectators.
A modified pulse sequence (the top six rows of Fig. 5) was used because it 
incorporates all the soft $\pi$
pulses in pairs, an arrangement well known to compensate pulse
imperfections [21].  The sign matrix corresponding to left-hand side of
the sequence shown in 
Fig. 5
is not a Hadamard matrix but is constructed by repeating sections of a 
Hadamard matrix.  The first six rows of the sign matrix are
\begin{eqnarray}
\begin{array}{ccccccccc}
I:\quad &+&+	&+	&+	&+	&+	&+	&+\\
S:\quad &+&+	&+	&+	&+	&+	&+	&+\\
R:\quad &+&-	&-	&-	&-	&+	&+	&+\\
Q:\quad 	&+	&+	&+	&-	&-	&-	&-	
&+\\
T:\quad 	&+	&+	&+	&+	&+	&+	&+	
&+\\
U:\quad 	&+	&-	&-	&-	&-	&+	&+	&+	
\end{array}
\end{eqnarray}
Note that no soft $\pi$ pulses are applied to spin $T$; row $T$ is 
therefore not orthogonal to rows $I$ or $S$.  This is allowed since
 we have assumed that $J_{IT}$ and $J_{ST}$ are negligible.

	The sequence for a controlled-not gate,  (3), employs evolution 
under
the $2I_zS_z$ operator for a period $(2n+1)/(2J_{IS})$, giving a multiplet
that is antiphase with respect to the $J_{IS}$ splitting, but which
requires a $\pi/2$ phase shift (the $I_z$
and $S_z$ terms) to convert from dispersion to absorption.  Sequences are
available [12] to implement the requisite $z$ rotation while returning the
spectator spins to their initial states.  For the purposes of illustration
we demonstrate only the evolution under the $IS$ coupling, obtaining the
phase shift by resetting the receiver reference phase when recording the
$I$ and $S$ responses.

The soft pulse scheme set out in Fig. 5 is designed to refocus vicinal
(three-bond) and long-range (four-bond) splittings.  In practice only the
vicinal couplings are well resolved for inosine in the rather viscous
solvent.

	The resulting spectra are shown in Fig. 6.  The soft $\pi$ pulses 
had a
duration of 64 ms and were flanked by 1 ms delays, giving an overall
duration of 528 ms for the sequence.   The geminal coupling between $I$ 
and
$S$ is 12.4 Hz, which requires $n=6$ in Eq.(4) to achieve the antiphase
condition while still accommodating all the soft $\pi$ pulses.  Although 
$I$
and $S$ show some effects of strong coupling (AB pattern) the antiphase
condition is quite clear, while the four spectator spins have chemical
shift and spin-spin couplings all refocused at the end of the sequence.
This implements the key step of the controlled-not logic gate.

	Systems with larger numbers of coupled spectator 
spins
are readily handled by extending the pulse patterns of Figs. 2, 3 or 5,
without increasing the overall duration of the sequence. The number of 
soft
$\pi$ pulses only increases linearly with the number of spectator spins.

To summarise -- there are not enough suitable spin-1/2 nuclear species to
be able to construct an entirely heteronuclear NMR quantum computer with
sufficient qubits to make it useful.  Hence any viable device must contain
extensive networks of coupled homonuclear spins, for example protons,
fluorine or carbon-13. 
This imposes constraints which rule out some ``efficient'' schemes
which are appropriate for heteronuclear systems [18,19] as the latter 
involve
simultaneous soft pulses on pairs of coupled spins, a procedure well known
to generate undesirable multi-spin coherences.  We have
demonstrated new sequences for the construction of a quantum logic gate
with homonuclear spins, where the spectator spins undergo no net evolution
for the duration of the gate.  When applied to the most realistic case
where each spin is coupled to a restricted set of neighbours (neglecting
long-range couplings) they have the important feature that the total
duration of the sequence does not increase as further spins are added to
the system.  Analogous considerations apply to systems made up of both
homonuclear and heteronuclear spins; the latter are refocused with
conventional hard $\pi$ pulses of negligible duration. We show experimental
results for a six qubit system: the six coupled protons in inosine.

} 

\begin{center}
Figure Captions
\end{center}

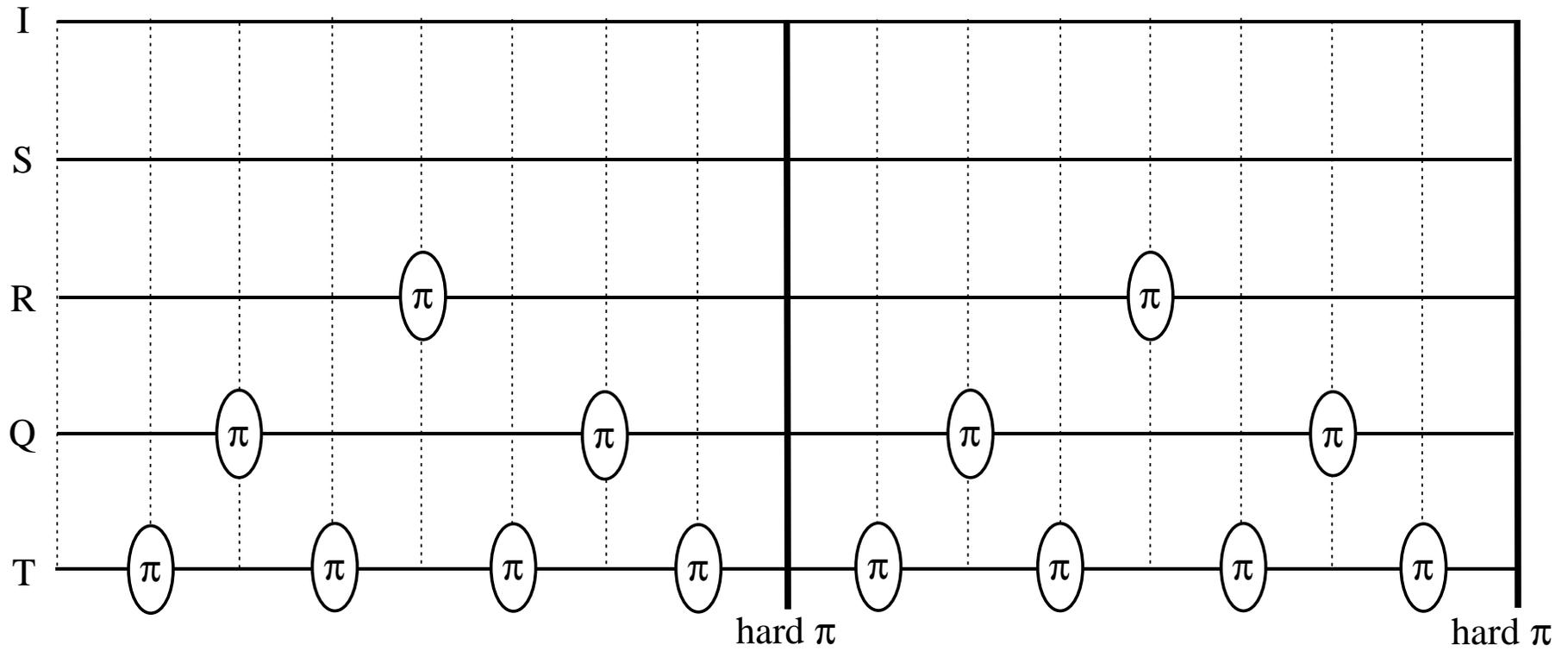
Fig. 1.   Refocusing scheme for a system of five homonuclear spins where
each pair of spins has an appreciable spin-spin coupling.   The ellipses 
represent
frequency-selective inversion pulses. The active spins
$I$ and $S$ evolve under the $2I_zS_z$ operator and the duration of the 
sequence
is set to $\tau = 1/(2J_{IS})$.  The spectator spins ($R$, $Q$, and $T$)
are returned
to their initial states at the end of the sequence.   Note that the 
introduction of an
additional spectator spin would necessitate a further stage of sixteen 
soft
$\pi$ pulses.

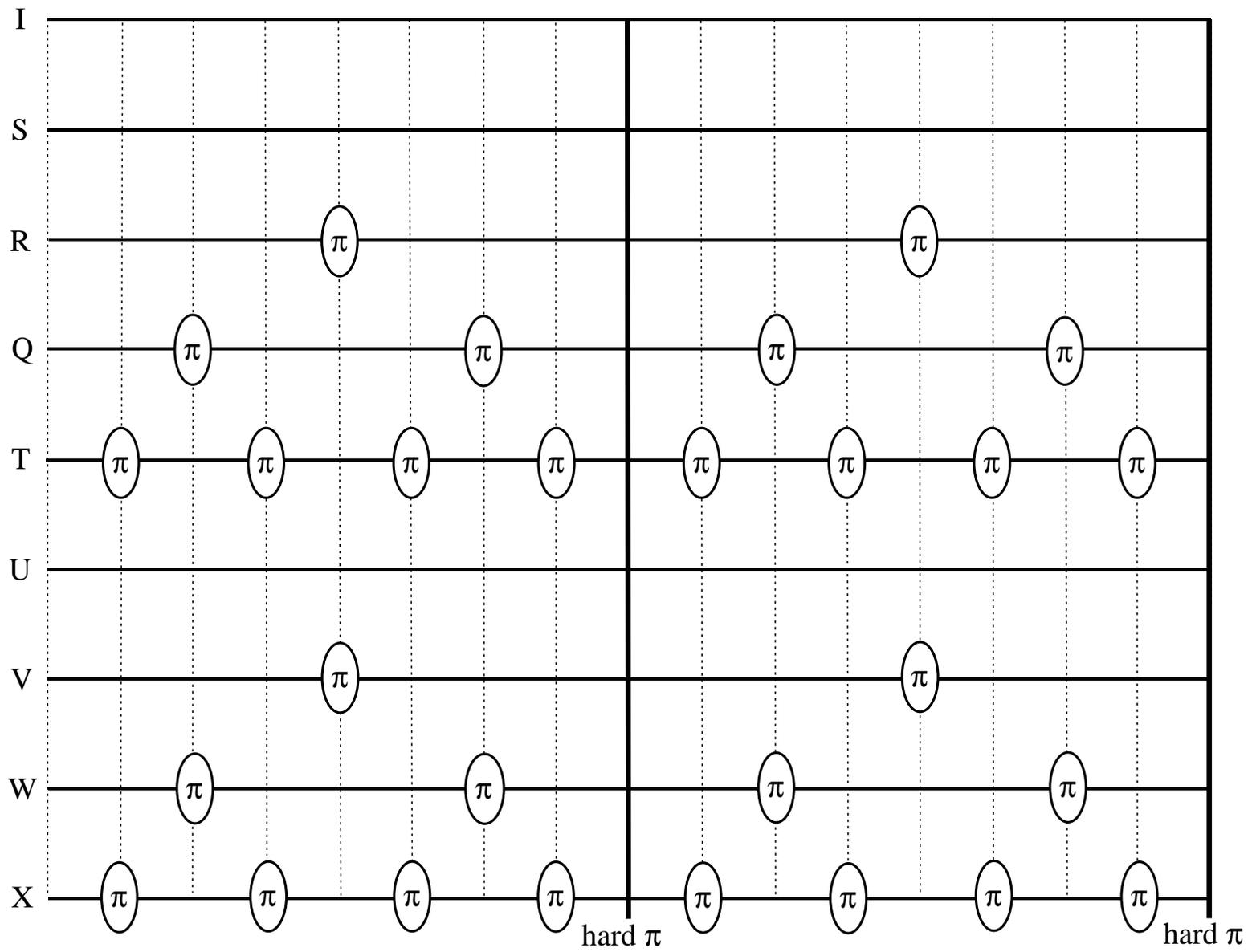
Fig. 2.  Refocusing scheme for a chain of nine homonuclear spins
for the case that spin-spin coupling over more than three chemical bonds
can be neglected.  $I$ and $S$ are the active spins evolving under the 
$2I_zS_z$
operator.  The (1-2-4) cascade pattern would be repeated as more spectator
spins are added.  Consequently the complexity increases essentially
linearly with the number of spectator spins.

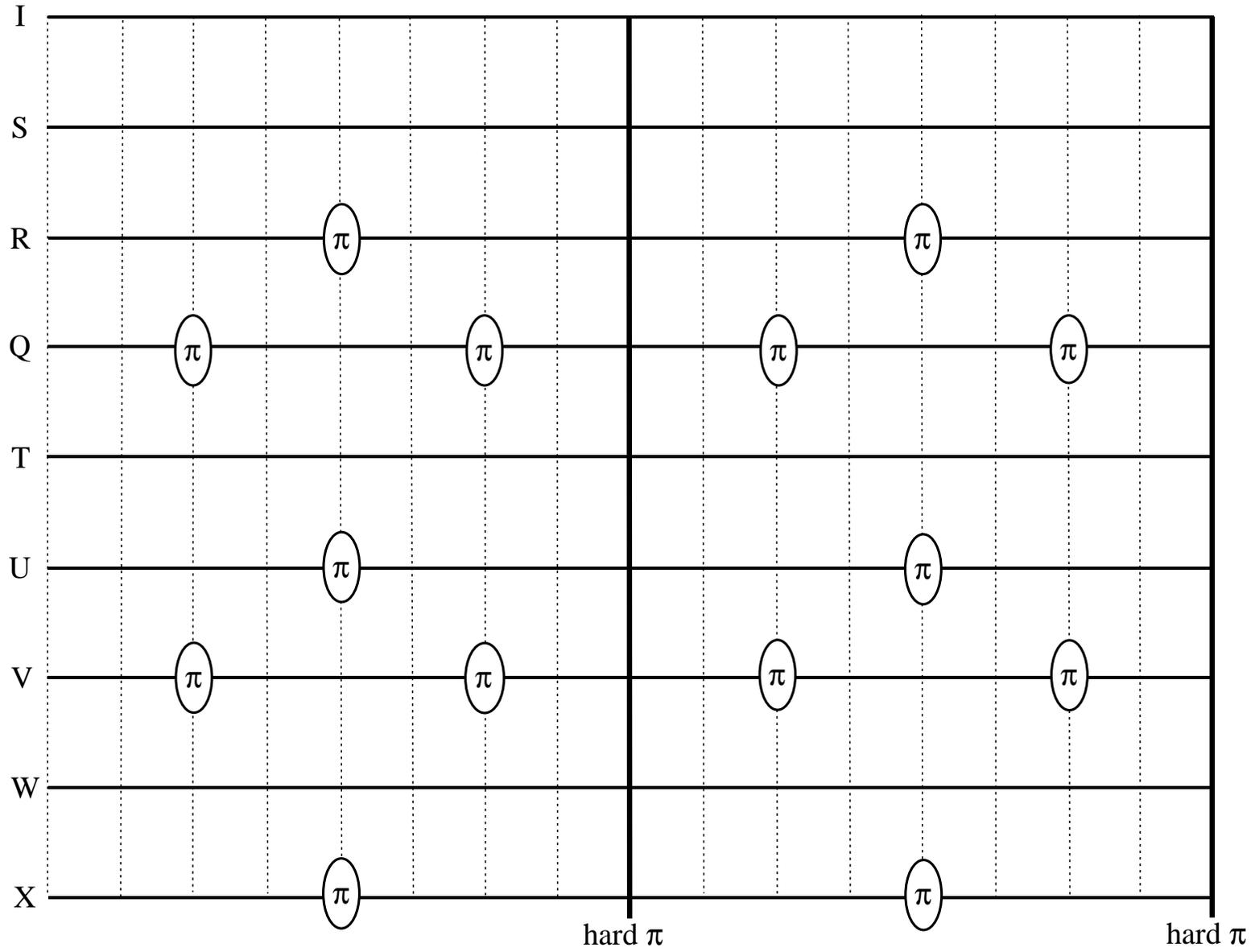
Fig. 3. Refocusing scheme similar to that shown in Fig. 2, except that
spin-spin coupling over more than two chemical bonds is neglected.  The
(1-2) cascade pattern may be repeated indefinitely as more spectator spins
are introduced without increasing the overall duration of the sequence.

Fig. 4. The six-spin system of inosine with active protons labelled
$I$ and $S$ and the spectators $R,Q,T$ and $U$.

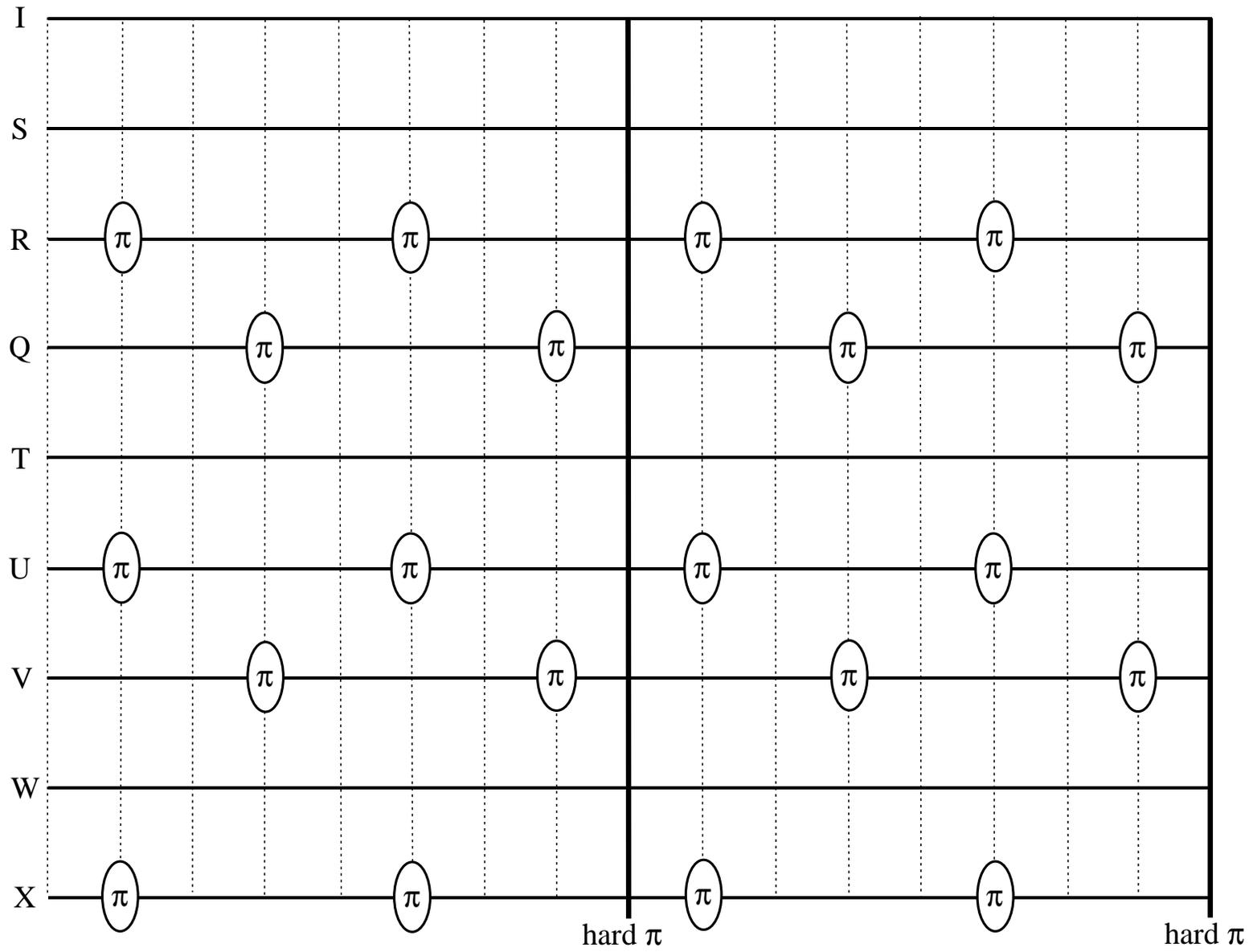
Fig. 5. A refocusing scheme equivalent to that shown in Fig. 3
incorporating soft $\pi$ pulses in pairs to compensate pulse 
imperfections.  The first six
rows of this sequence were used to obtain the experimental results shown 
in
Fig. 6.

Fig. 6.  (a) Part of the conventional 600 MHz spectrum of the six protons
of inosine.  (b) The spectrum obtained after the pulse sequences of Fig. 5,
showing the $I$ and $S$ responses in antiphase dispersion with respect to
$J_{IS} = 12.4 Hz$.  (c) Individual multiplets expanded in frequency 3.6
times, with a $\pi/2$ phase shift applied to $I$ and $S$ to restore the
absorption mode.  Note that the spectators ($R$, $Q$, $T$ and $U$) are
essentially unchanged at the end of the sequence, apart from some minor
attentuation through spin-spin relaxation.

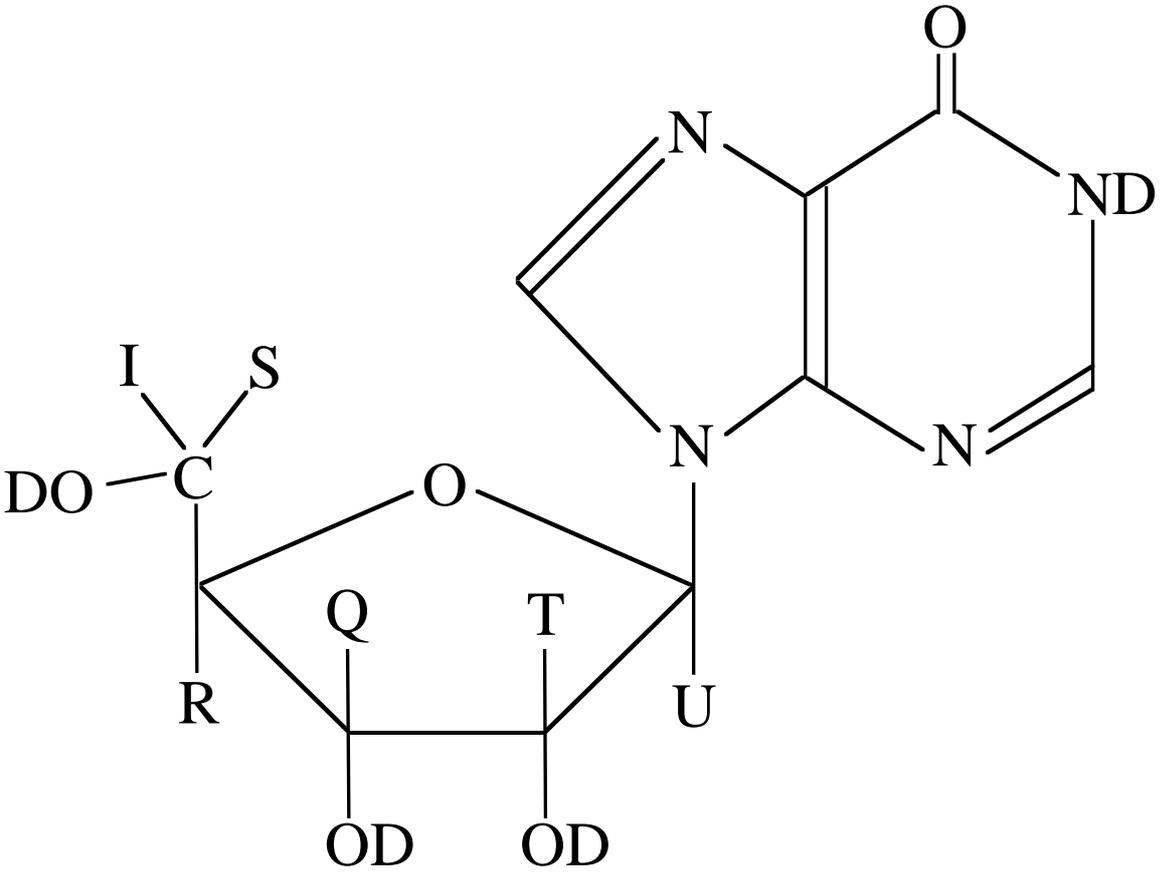
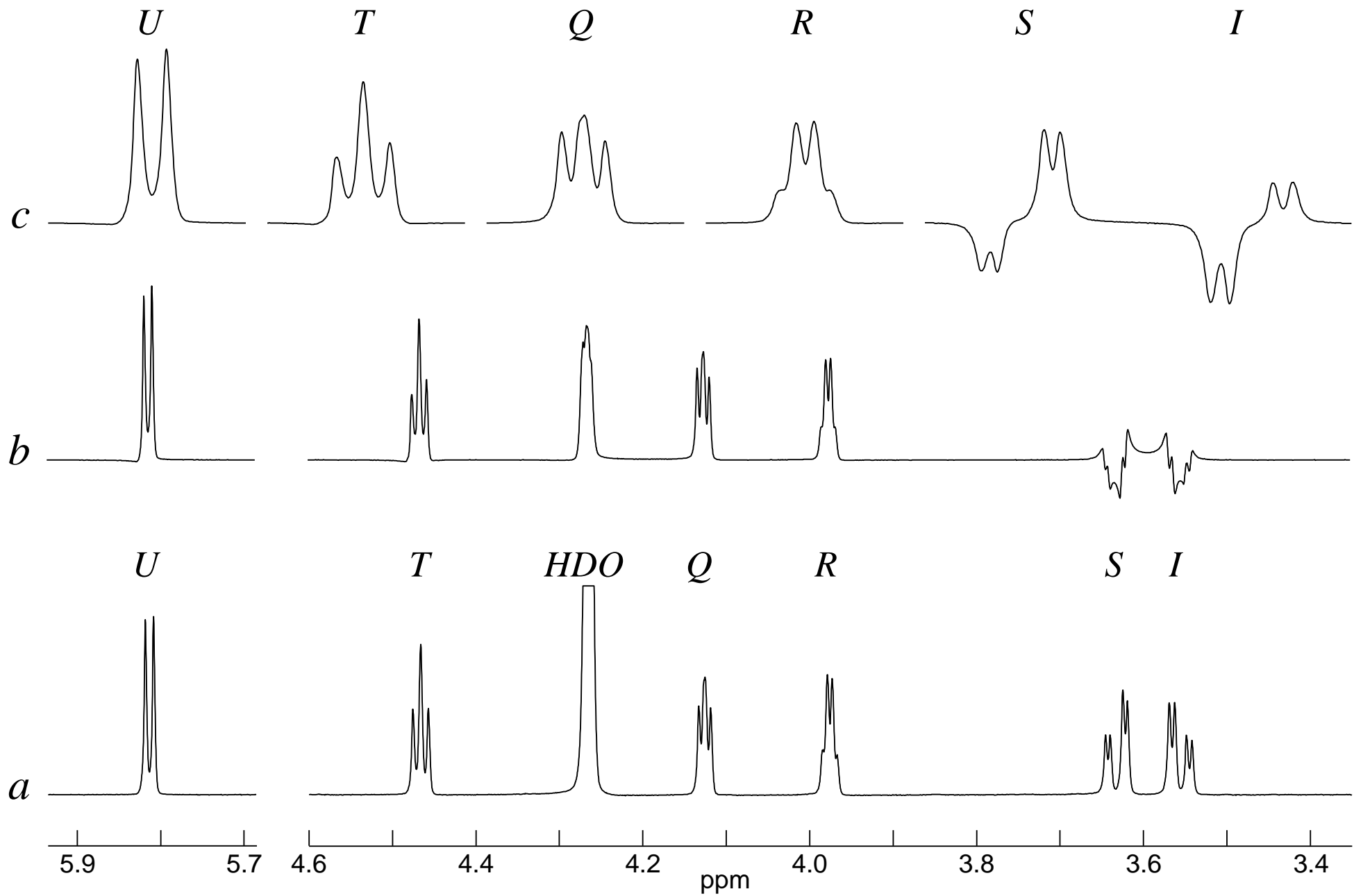
\end{document}